\documentstyle[twocolumn,epsf]{jpsj}
\catcode`\@=11
\newcount\@arga
\newcount\@argb
\newcount\@argc
\newcount\@ctmpa
\newcount\@ctmpb
\newcount\@ctmpc
\newcount\@ctmpd
\newcount\@ctmpe
\newdimen\@darg
\newdimen\@bblen
\newif\if@bbllx
\newif\if@bblly
\newif\if@bburx
\newif\if@bbury
\newif\if@height
\newif\if@width
\newif\if@scale
\newif\ifno@bb
\newif\ifepsfdraft
\epsfdraftfalse
\def\@setpsfile#1{
                \typeout{epsf:[#1]}
                \def\@psfile{#1}
}
\def\@setpsheight#1{
                \@heighttrue
                \@darg=#1\relax
                \edef\@psheight{\number\@darg}
}
\def\@setpswidth#1{
                \@widthtrue
                \@darg=#1\relax
                \edef\@pswidth{\number\@darg}
}
\def\@setpsscale#1{
                \@scaletrue
                \def\@pshscale{#1}
                \def\@psvscale{#1}
                \@bblen=#1pt\relax
                \@bblen=1000\@bblen
                \def\@texhscale{\expandafter\remove@dim\the\@bblen}
                \let\@texvscale=\@texhscale
}
\def\@setpshscale#1{
                \@scaletrue
                \def\@pshscale{#1}
                \@bblen=#1pt\relax
                \@bblen=1000\@bblen
                \def\@texhscale{\expandafter\remove@dim\the\@bblen}
}

\def\@setpsvscale#1{
                \@scaletrue
                \def\@psvscale{#1}
                \@bblen=#1pt\relax
                \@bblen=1000\@bblen
                \def\@texvscale{\expandafter\remove@dim\the\@bblen}
}

\def\@setparms#1=#2,{\@nameuse{@setps#1}{#2}}
\def\ps@init@parms{
                \@heightfalse \@widthfalse
                \no@bbfalse
                \def\@psbbllx{}\def\@psbblly{}
                \def\@psbburx{}\def\@psbbury{}
                \def\@psheight{}\def\@pswidth{}
                \def\@pshscale{1}\def\@psvscale{1}
                \def\@texhscale{1000}\def\@texvscale{1000}
                \def\@psfile{}
                \def\@sc{}
}
\def\parse@ps@parms#1{
                \@for\@epsfile:=#1\do
                   {\expandafter\@setparms\@epsfile,}}
\newif\ifnot@eof
\newread\ps@stream
\def\bb@search{
        \openin\ps@stream=\@psfile
        \no@bbtrue
        \not@eoftrue
        \catcode`\%=12\relax
        \ifeof\ps@stream\typeout{epsf: File not found}\fi
        \loop
                \read\ps@stream to \line@in
                \global\toks200=\expandafter{\line@in}\relax
                \ifeof\ps@stream \not@eoffalse \fi
                \@bbtest{\toks200}\relax
                \if@bbmatch\not@eoffalse\expandafter\bb@cull\the\toks200\fi
        \ifnot@eof \repeat
        \catcode`\%=14
}       

\catcode`\%=12
\newif\if@bbmatch
\def\@bbtest#1{\expandafter\@a@\the#1
\long\def\@a@#1
        \ifx\@bbtest#2\@bbmatchfalse\else\@bbmatchtrue\fi}
\def\bb@cull 
        \@ifnextchar\space{\@latexbug}{\bb@extract}}
\def\bb@extract #1 #2 #3 #4 {
        \message{BoundingBox: (#1bp,#2bp)--(#3bp,#4bp)}
        \@darg=#1 bp\edef\@psbbllx{\number\@darg}
        \@darg=#2 bp\edef\@psbblly{\number\@darg}
        \@darg=#3 bp\edef\@psbburx{\number\@darg}
        \@darg=#4 bp\edef\@psbbury{\number\@darg}
        \no@bbfalse
}
\catcode`\%=14

\def\compute@bb{
                \bb@search
                \ifno@bb \typeout{epsf: No BoundingBox}
                \stop
                \else
                \@arga=\@psbburx
                \advance\@arga by -\@psbbllx
                \edef\@bbw{\number\@arga}
                \@arga=\@psbbury
                \advance\@arga by -\@psbblly
                \edef\@bbh{\number\@arga}
                \fi
}
\def\in@hundreds#1#2#3{\@argb=#2 \@argc=#3
                     \@ctmpa=\@argb     
                     \divide\@ctmpa by \@argc
                     \@ctmpb=\@ctmpa
                     \multiply\@ctmpb by \@argc
                     \advance\@argb by -\@ctmpb
                     \multiply\@argb by 10
                     \@ctmpb=\@argb     
                     \divide\@ctmpb by \@argc
                     \@ctmpc=\@ctmpb
                     \multiply\@ctmpc by \@argc
                     \advance\@argb by -\@ctmpc
                     \multiply\@argb by 10
                     \@ctmpc=\@argb     
                     \divide\@ctmpc by \@argc
                     \@arga=#1\@ctmpe=0
                     \@ctmpd=\@arga
                        \multiply\@ctmpd by \@ctmpa
                        \advance\@ctmpe by \@ctmpd
                     \@ctmpd=\@arga
                        \divide\@ctmpd by 10
                        \multiply\@ctmpd by \@ctmpb
                        \advance\@ctmpe by \@ctmpd
                     \@ctmpd=\@arga
                        \divide\@ctmpd by 100
                        \multiply\@ctmpd by \@ctmpc
                        \advance\@ctmpe by \@ctmpd
                     \edef\@result{\number\@ctmpe}
}
\def\compute@wfromh{
                \in@hundreds{\@psheight}{\@bbw}{\@bbh}
                \edef\@pswidth{\@result}
}
\def\compute@hfromw{
                \in@hundreds{\@pswidth}{\@bbh}{\@bbw}
                \edef\@psheight{\@result}
}
\def\compute@handw{
        \if@height 
                \if@width
                \else
                        \compute@wfromh
                \fi
        \else 
                \if@width
                        \compute@hfromw
                \else
                        \if@scale
                                \in@hundreds{\@texvscale}{\@bbh}{1000}
                                \let\@bbh=\@result
                                \in@hundreds{\@texhscale}{\@bbw}{1000}
                                \let\@bbw=\@result
                        \fi
                                \edef\@psheight{\@bbh}
                                \edef\@pswidth{\@bbw}
                \fi
        \fi
}
{\catcode`\p=12\catcode`\t=12
\gdef\remove@dim#1.#2pt{#1}}

\def\compute@sizes{
        \compute@bb
        \compute@handw
}
\def\epsfile#1{
        \ps@init@parms
        \parse@ps@parms{#1}
        \compute@sizes
        \@arga=\@psheight
        \divide\@arga by 65536
        \edef\@psvsize{\number\@arga}
        \@arga=\@pswidth
        \divide\@arga by 65536
        \edef\@pshsize{\number\@arga}
        \message{=>(\@pshsize bp,\@psvsize bp)}
        \leavevmode
        \vbox to \@psheight true sp{
                \hbox to \@pswidth true sp{
                \ifepsfdraft\hss\@psfile\hss\else
                \if@height 
                        \if@width
                                \special{epsfile=\@psfile \space 
                                hsize=\@pshsize \space
                                vsize=\@psvsize \space}
                        \else
                                \special{epsfile=\@psfile \space 
                                vsize=\@psvsize \space}
                        \fi
                \else 
                        \if@width
                                \special{epsfile=\@psfile \space 
                                hsize=\@pshsize \space}
                        \else
                                \if@scale
                                        \special{epsfile=\@psfile \space
                                        vscale=\@psvscale \space
                                        hscale=\@pshscale \space}
                                \else
                                        \special{epsfile=\@psfile \space}
                                \fi
                        \fi
                \fi
                \hfil\fi
                }
        \vfil
        }
}

\catcode`\@=12
\begin{document}
\renewcommand{\thefootnote}{\fnsymbol{footnote}}
\def\runtitle{
Scaling Properties of Spin Ladder with Impurities
}
\title
{
Scaling Properties of Antiferromagnetic Transition in Coupled
  Spin Ladder Systems Doped with Nonmagnetic Impurities
\footnote{to appear in J. Phys. Soc. Jpn.{\bf 66}(1997) No.3}
}

\def\runauthor{
Masatoshi I{\sc mada} and Youichirou I{\sc ino}
}

\author{%
Masatoshi I{\sc mada} and Youichirou I{\sc ino}
}

\inst{
  Institute for Solid State Physics,\\
  University of Tokyo, Roppongi 7-22-1,\\
  Minato-ku, Tokyo 106
}

\recdate
{
January 10, 1997
}

\abst{
We study effects of interladder coupling on critical magnetic properties of
spin ladder systems doped with small concentrations of nonmagnetic
impurities, using the scaling theory together with 
quantum Monte Carlo (QMC) calculations.  Scaling properties in a wide region in
the parameter space of the impurity concentration $x$ and the
interladder coupling are governed by the quantum critical point (QCP) of the 
undoped system for the transition between antiferromagnetically ordered
and spin-gapped phases.  This multi-dimensional and strong-coupling 
region has 
characteristic power-law dependences on $x$ for magnetic properties
such as the N\'eel temperature.  The relevance of this criticality for
understanding experimental results of ladder compounds is stressed.  
}
\kword
{spin ladder, scaling theory, quantum critical point,
nonmagnetic impurity, antiferromagnetic order
}
 
\maketitle
\sloppy

Mott insulators on several geometrically designed structures of
lattices are known to have the spin excitation gap from the spin singlet
ground state\cite{1}.  
Several compounds  such as SrCu$_2$O$_3$\cite{2} and
(VO)$_2$P$_2$O$_7$\cite{3} indeed have the gap which may be attributed to the
underlying ladder structure of Cu$_2$O$_3$ or V$_2$O$_3$ network.  The 
spin-1/2 Heisenberg model on a ladder given by the Hamiltonian 
\begin{equation}
{\cal H} = \sum_{\langle i,j\rangle}J_{ij}{\bf S}_i\cdot{\bf S}_j \label{1}
\end{equation}
with the exchange interaction between nearest-neighbor
pairs $\langle i,j \rangle$ describes low-energy spin
excitations.  The spin gap $\Delta_s$ is as large as 0.5$J$ for the
uniform strength of $J=J_{ij}$\cite{4}.    

Recent experiments have revealed that the 
spin ladder compound SrCu$_2$O$_3$ shows surprisingly sensitive
dependence on small concentration of Zn substitution for Cu sites\cite{5,6}.
Even at the doping concentration of Zn as low as $x=0.01$, the system 
shows anomalies in the
susceptibility and specific heat at the temperature $T~\sim$~3K.
The $x$ dependence suggests that the compound changes 
from a spin-gapped insulator at $x=0$ to an insulator with an 
antiferromagnetic long-range order (AFLRO) below the temperature of
the 
anomaly, $T_N$, which quickly increases
with increasing $x$.  In addition, above $T_N$, it shows a
remarkable Curie-like behavior $\chi=C/T$ for the uniform susceptibility
and a $T$-linear specific heat $C_p$.    
The sensitive dependence on $x$ and the appearance of the
AFLRO have a similarity to the case of a spin-Peierls 
compound CuGeO$_3$\cite{7}.  

Since the Zn substitution plays a role to deplete the spin-1/2 local
moment on a half-filled Cu $d_{x^2-y^2}$ orbital, the ladder model (1)
with a dilute depletion of sites has been extensively studied
theoretically.  Variational Monte Carlo, as well as the exact
diagonalization results show that a nonmagnetic impurity induces the
formation of a static spin-1/2 moment with a substantial enhancement of 
antiferromagnetic correlations (AFC) around the impurity\cite{8}.  
Quantum Monte Carlo (QMC)
results of a single ladder have supported this result of the 
enhancement\cite{9}.  This
enhancement has also been suggested in other theoretical approaches for models with a spin gap\cite{10,11,12}.
The QMC results\cite{9} on ladders have further reproduced  the 
Curie law, $\chi=C/T$ and the linear-$T$ specific heat $C_p=\gamma T$ with quantitative agreements of $C$ 
and $\gamma$ with the experimental results, when the nonmagnetic
impurity concentration $x$ exceeds a rather sharp crossover concentration $x_c
\simeq 0.04$-0.05.  
However, below $x_c$,
the local moments induced around the depleted sites are so weakly
coupled that they do not explain the experimental results for
$\gamma$ and the appearance of AFLRO.  
This is due to the fact that the effective coupling between the induced 
spin-1/2 moments at distance $r$ decreases exponentially as
${\rm exp}[-r/\xi_0]$ and this characteristic energy scale for the
typical distance between neighboring induced moments becomes far below 
the temperature in experiments for realistic choices of
the AFC length $\xi_0$ of a single ladder.  On the contrary, the experimental
results must be
the consequence of strongly and
antiferromagnetically coupled moments.  

\begin{figure}[btp]
\hfil
\epsfile{file=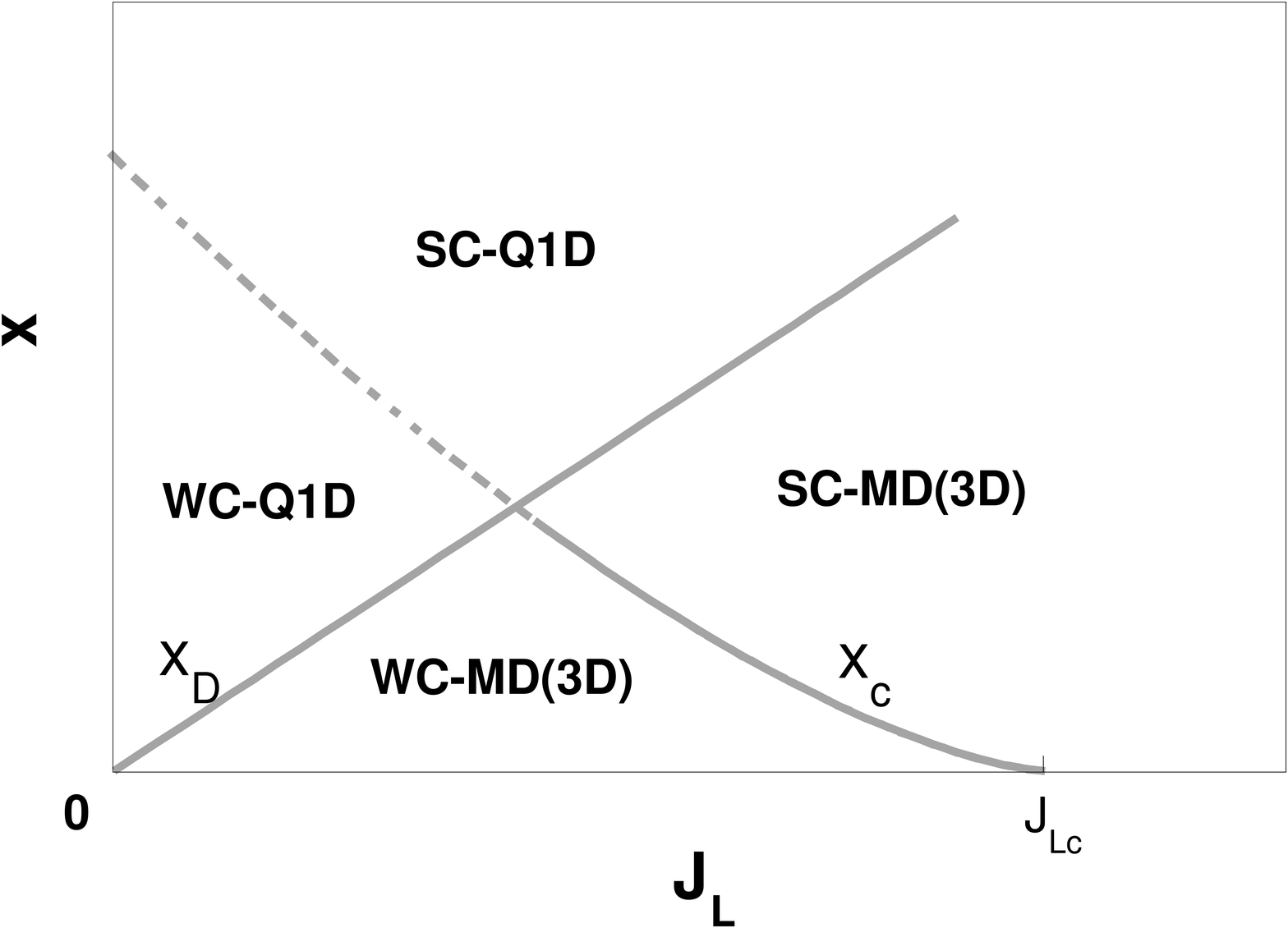,scale=0.075}
\hfil
\caption{Schematic phase diagram for the scaling properties in
  the plane of interladder coupling $J_L$ and nonmagnetic impurity
  concentration $x$.  The quasi-one-dimensional (Q1D) and
  multi-dimensional (MD) regions are separated by the solid curve
   $x_D$ while the strong coupling (SC) and weak coupling (WC) regions are 
separated by 
  the curve $x_c$.}
\label{fig1}
\end{figure}

In this letter we study effects of interladder coupling to understand
this puzzling feature.  We show existence of four different regions
 of the critical behavior separated by two crossover lines in the
parameter 
space of the
interladder coupling $J_L$ and 
$x$.  Previously employed quasi-one-dimensional (Q1D)
and weak-coupling (WC)  
approaches from the single ladder in the literature are valid in one
of the four regions, whereas we stress that a different region 
among the four is relevant for understanding experimental indications.  
This region is governed by the universality of the QCP for undoped systems, where the scaling properties are
characterized by strong coupling (SC) of the induced moments and the
multi-dimensional (MD) criticality.  We discuss the scaling properties of
this antiferromagnetic transition.  QMC results on
coupled ladders as well as experimental indications are compared with
the scaling predictions.  We summarize our basic phase diagram for
different scaling properties in Fig.\ref{fig1}.  Detailed definitions of WC,
SC, Q1D and 3D (three-dimensional) or MD are clarified in the
discussions below. 

First we estimate the scaling properties of the antiferromagnetic
transition and the upper bound for the N\'eel temperature in the Q1D 
region of weakly coupled (WC) ladders.  This region is realized when
the interladder coupling is finite but sufficiently weak.  In this region,  
with decreasing temperature,
$\xi_0$ within and along the ladder becomes much longer than
$\xi_L$ for the interladder direction.  
The effective exchange coupling between two local moments induced 
at two depleted sites each depends exponentially on the distance
between these depleted sites, $r$, as 
$J(r) \simeq J_0 e^{-r/\xi_0}$
with $J_0$ being the nearest-neighbor exchange coupling of the 
uniform 
ladder.    
Here we neglect logarithmic corrections.  
When the impurities are doped
randomly, the distribution function of the distance between two
neighboring depleted sites follows the Poisson-type distribution
$P_r(r) = \frac{r}{l^2} e^{-r/l}$
for the concentration of depleted sites, $x=1/2l$.  Here, the length
is scaled by the lattice constant of the ladder.  From
the above relations the distribution function for the amplitude
of the exchange coupling between the moments at neighboring depleted
sites follows
\begin{equation}
P_J(J) \sim - \frac{\xi_0^2}{l^2} ({\rm
  ln}\frac{J}{J_0})\frac{1}{J}(\frac{J}{J_0})^{\frac{\xi_0}{l}}.
\label{4}
\end{equation}
When $T$ is lowered, only $J$
above the scale of $T$ can contribute to the growth of
AFC.  The ratio of such coupling, $R$, among
all the neighboring exchange between depleted sites is estimated as 
\begin{equation}
R =  \int_{J_{th}}^{J_0}P_J(J)dJ
  =  1 + \frac{\xi_0}{l}(\frac{J_{th}}{J_0})^{\frac{\xi_0}{l}}\left({\rm
   ln}\frac{J_{th}}{J_0}- \frac{l}{\xi_0}\right),  \label{5}
\end{equation}
where we take $J_{th}\sim T$.  
When 
$x$ is low enough so that $l\gg\xi_0$ is
satisfied,
$R\simeq 1-({J_{th}}/{J_0})^{\xi_0/l}$ 
follows.  Therefore the AFC along the ladder 
is developed over many depleted sites only for not too small $R$ and
hence only at
\begin{equation}
T<T_{cr1}=J_0 e^{-\frac{a}{x}}  \label{7}
\end{equation}
with
$a=\frac{1}{2\xi_0}\simeq\frac{\Delta_s}{J_0}$. 
  As we have neglected the quantum fluctuation for 
progressive singlet  formation on the stronger bonds as is frequently discussed
in the random exchange antiferromagent\cite{9,10}, $T_{cr1}$ 
based on the above argument of the percolation threshold is certainly an
upper bound temperature for the growth of a long correlation length
within a ladder.  

When we
consider the interladder couplings $J_{L1}$ and $J_{L2}$, in the other
two directions, the 
necessary condition for the growth of 3D
correlation over many depleted sites in the WC region is similarly given from
\begin{equation}
R\simeq 1-(\frac{T}{J_0})^{\frac{\xi_0}{l}}
(\frac{T}{J_{L1}})^{\frac{\xi_{L1}}{l}}
(\frac{T}{J_{L2}})^{\frac{\xi_{L2}}{l}}  \label{8}
\end{equation}
with $2l=1/x^{1/3}$   
as
\begin{equation}
T< T_{cr3} = J_0{\rm exp} \left[-\frac{\Delta_s}{(J_0J_{L1}J_{L2}x)^{1/3}}\right] 
\label{10}
\end{equation}
at small enough $x$.  
The upper bound of $T_N$ in the $3D$ region is given by 
$T_{cr3}$.

From (\ref{7}) and (\ref{10}), it turns out that the Q1D treatment is
always invalid at small $x$.  The boundary to separate the Q1D and 
the 3D region is given by 
\begin{equation}
x_D \sim \frac{\sqrt{J_{L1}J_{L2}}}{J_0}.  \label{13}
\end{equation}
Below $x_D$, three dimensional correlations are developed before 
the correlation grows only along the ladder.  Equations (\ref{7}) and
(\ref{10}) show that, in the region of weakly coupled local moments, the 
upper bound of $T_N$ is exponentially small with an essential
singular dependence on $x$.   
In the WC region, the upperbound of the equal-time spin structure factor 
$\lim_{N\rightarrow\infty} S(Q)/N \equiv \sum_r \langle s_0^zs_r^z\rangle e^{iQr}$ at the staggered wave
vector $Q$ should be scaled as 
\begin{equation}
\lim_{N\rightarrow\infty}S(Q)/N= \langle M_s\rangle^2/3 \propto x^2, \label{30} 
\end{equation}
because only the spin-1/2 local moments induced around the depleted
sites contribute to the order parameter.  Here $\langle
M_s\rangle$ is the staggered magnetization.  

A qualitatively different $x$ dependence of $T_N$ is derived in the
SC region which is governed by the QCP at $J_{L1}=J_{L1c}$ and $J_{L2}=J_{L2c}$.  This critical point is
defined as the interladder coupling beyond which the AFLRO is stabilized at
$T=0$ 
even without the nonmagnetic impurities.
To discuss the scaling properties of the antiferromagnetic transition
in the SC region, we take $\delta=J_{Lc}-J_{L}, J_L\equiv
J_{L1}=J_{L2}$ and $J_{Lc}\equiv J_{L1c}=J_{L2c}$ for simplicity.

At $x=0$, the correlation length exponent $\nu$ and the staggered
magnetization exponent $\beta$ are defined as 
$\xi \propto \delta^{-\nu}$
and 
$\langle M_s\rangle \propto |\delta|^{\beta}$ for small $|\delta|$.    
The N\'eel temperature $T_N$ for $\delta<0$ is scaled from the
critical point as 
$T_N \propto (-\delta)^{\nu z}$   
where the dynamical exponent $z$ is introduced.  

When $x$ is finite
but $\delta=0$, we can derive scaling exponents for $x$ dependence
because $x$ is directly related with the inverse cube of the length
scale.  For example $T_N$ should be characterized by the temperature
$T_{cr}^*$ below which the averaged correlation length $\xi$ exceeds the
mean impurity distance $x^{-1/3}$.  Below $T_{cr}^*$, 
the local moments induced around the depleted sites become strongly
correlated antiferromagnetically through the overlap of the correlated region of the radius
$\xi$.  With
lowering temperature, $\xi\sim 1/T^z$ is expected.  This gives the scaling   
\begin{equation}
T_N \propto x^{p}  \label{17}
\end{equation}
with $p=z/3$ in three dimensions.     
Similarly $\langle M_s\rangle$ at $T=0$ 
is scaled as 
\begin{equation}
\langle M_s\rangle 
\propto x^{q} \label{19}
\end{equation}
with $q=\beta/d\nu$ in $d$ spatial dimensions for $d \geq 2$.  
These scaling properties at $\delta=0$ govern the SC region.
As compared to the WC region which shows the essentially singular
dependence on $x$, characteristic scales all have power-law dependence 
on $x$.    
  The crossover boundary between the WC and SC regions is given by 
\begin{equation}
x_c = \delta^{u}  \label{20}
\end{equation}
with $u=d\nu$.  
Therefore when we fix $\delta$ and increase $x$, physical quantities show 
a crossover
from the scaling in the WC region to that in the SC region at this
boundary.  

The universality class of the quantum transition at $x=0$ as a
function of $\delta$ in $d$ dimensions is not exactly known.  However, 
there exists growing support for the equivalence to the $d+1$-dimensional 
nonlinear sigma model\cite{13,14}.   The
dynamical exponent $z$ is then given by $z=1$.
The universality class of the four-dimensional nonlinear sigma model is characterized 
by the mean field exponents $\nu=0.5$ and $\beta=0.5$.  
Then for the exponents in the SC region, 
we obtain $2p=q=2/3$ and $u=3/2$ for the 3D case.  For comparison, we note
that, in 2D, the long-range order may appear only
at $T=0$, where $q\sim 0.26$ and $u\sim 1.4$ are obtained because
$\nu\simeq 0.70$ and $\beta\simeq 0.36$ are expected\cite{30}.     
It should be noted that the boundary $x_c$ may be rather small because 
$u>1$ and 
$x_c$ is extrapolated to around 0.04 even for the single 1D ladder  
 as clarified in the QMC results\cite{9}.  

We next discuss quantitative aspects of the scaling properties by
using the QMC results.    
In this letter, we show the QMC results 
of coupled ladders  
 for the configuration 
of the 
two-dimensional square lattice.  
This structure has two basic differences from that of real
compounds such as SrCu$_2$O$_3$.  One is that
the three-dimensionality is ignored so that we cannot discuss scaling
properties of $T_N$ directly.  The other point is that the
experimental lattice structure has a certain frustration on the ladder plane while it is not
considered here.   However, even
with these drawbacks of the present model, it is 
still useful to understand the four different criticalities in the
WC-Q1D, WC-MD, SC-Q1D and SC-MD regions with the  
robustness of the SC-MD region in the simplest 2D case.  
More detailed analyses with a 3D realistic lattice 
structure will be discussed elsewhere.  QMC
calculations have been performed by using the world line algorithm with a
cluster updating method\cite{15} and a continuous time
procedure\cite{16}.
The system size is $N$=$L_x\times L_y$
with $L_y$ being the ladder direction.  Several choices of linear
dimensions are taken for the extrapolation between $L_y=$32 and 128 for
$L_y/L_x=4,6$ and 8.

\begin{figure}[btp]
\hfil
\epsfile{file=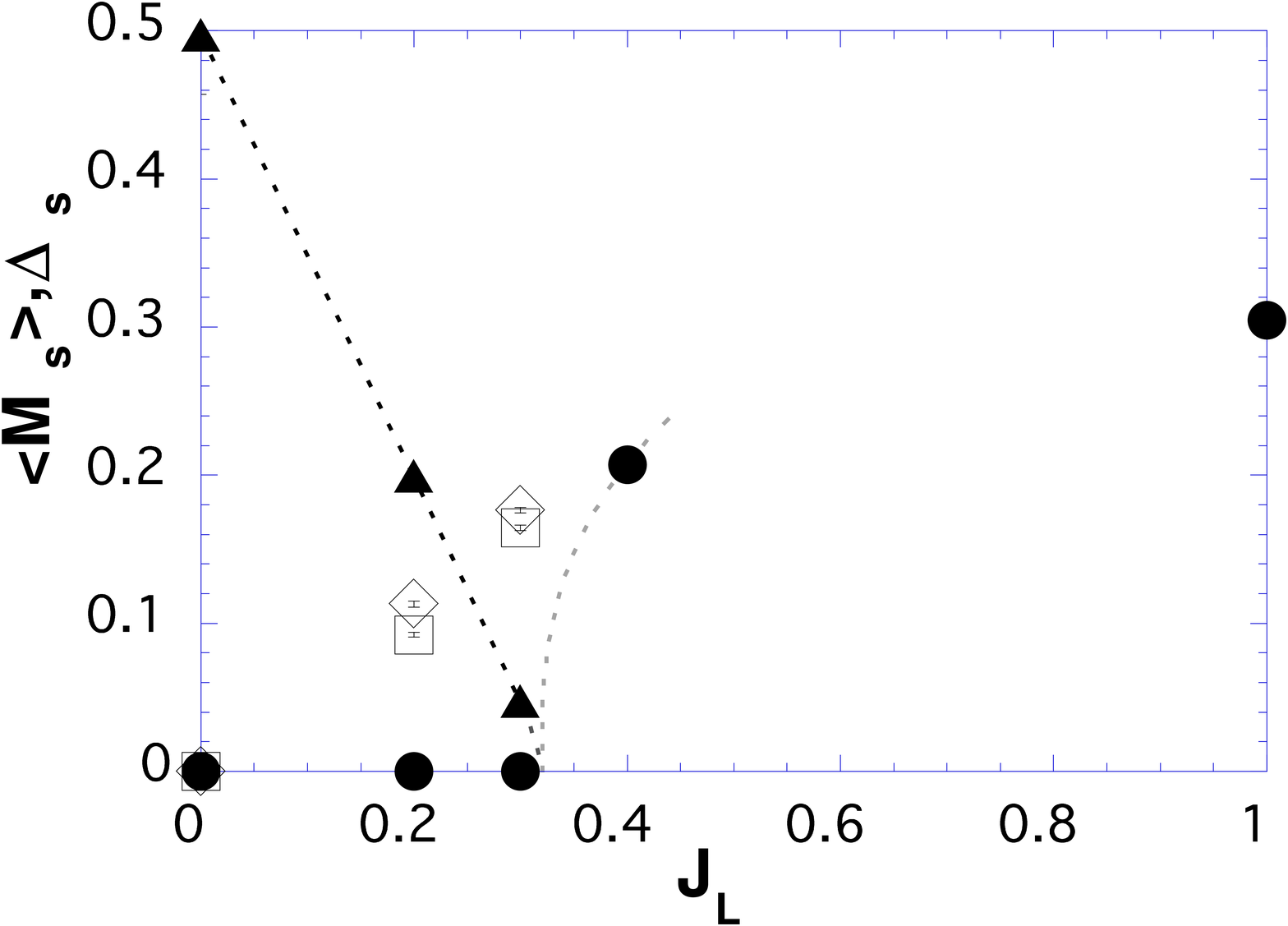,scale=0.095}
\hfil
\caption{The spin gap $\Delta_s$ (triangle) and the staggered 
magnetization $\langle M_s\rangle$ (circle) as  
a function of $J_L$  scaled by $J_0$ at $T=0$ and $x=0$.  Squares and
diamonds show $\langle M_s\rangle$ for $T=0$ at $x=0.0156$ and 0.0208, 
respectively.  The dotted lines are guides for the eyes.} 
\label{fig2}
\end{figure}

\begin{figure}[btp]
\hfil
\epsfile{file=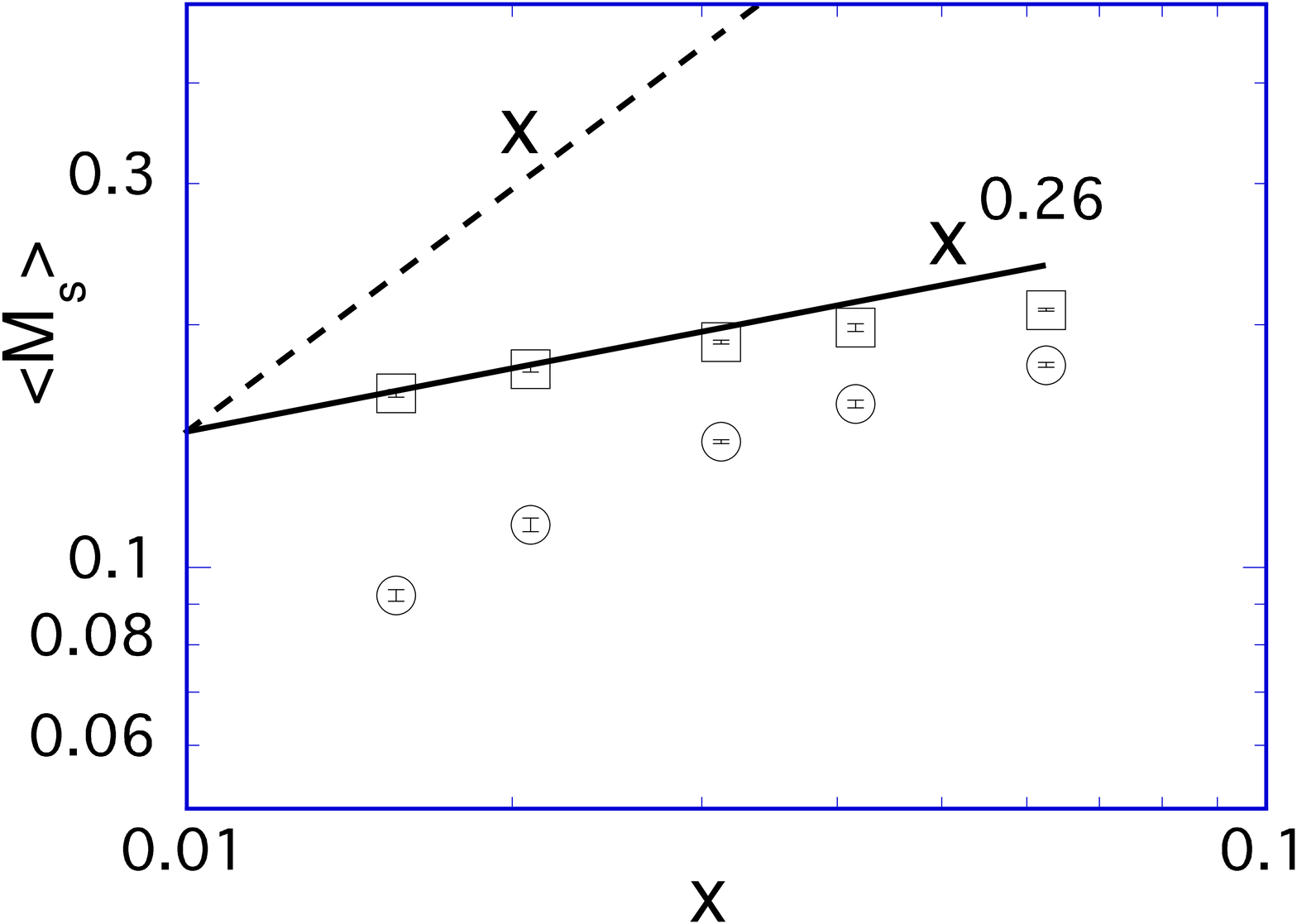,scale=0.1}
\hfil
\caption{$x$ dependence of $\langle M_s\rangle$ for $J_L=0.2$
  (circle) and
  0.3 (square) at $T=0$.  Solid (broken) lines show the slopes for the 
  scalings of the SC (WC) region.} 
\label{fig3}
\end{figure}

We first discuss quantitative aspects at 
$x=0$.  At $x=0$, the QCP for the antiferromagnetic 
transition is estimated as $J_{Lc}=0.32\pm 0.02$ as illustrated in
Fig.\ref{fig2}.  For $J_L>J_{Lc}$, the AFLRO 
appears at $T=0$ and $x=0$.  Two methods have been employed to
estimate $J_{Lc}$.  One is to measure the $J_L$ dependence of the spin 
gap for $J_L<J_{Lc}$.  For finite size lattices, the spin gap
$\Delta_s$ is determined from the fitting of the uniform
susceptibility $\chi_0$ to the form 
\begin{equation}
\chi_0(T) = A {\rm exp}[-\Delta_s/T].  \label{24}
\end{equation}
The limit $N\rightarrow\infty$ is taken in the scaling of
$\Delta_s$ with $1/N^2$.  Figure \ref{fig3} shows the extrapolated values
of $\Delta_s$.  
In 
the other method, $J_L$ dependence of $\langle M_s\rangle$ has been examined in
the region of $J_L >J_{Lc}$.  The critical point is consistent with
$J_{Lc}=0.32\pm 0.02$ in both of the analyses. 

When $x\not=0$, it is expected that the AFLRO appears in all the
regions of 
$J_L>0$, if $x$ is not too large.
However, scaling properties for the $x$ dependence may be drastically
different for different regions as we mentioned above.  Here, we show
$x$ dependence of $\langle M_s\rangle$ at $J_L=0.2$ and 0.3. 
In the following, we show results for regular distribution of
impurities.  The results for the random distribution show
qualitatively similar behavior although the AFLRO is 
quantitatively suppressed. 
Detailed analyses will be discussed elsewhere.   
 Note that at $x=0$,
the gap extrapolated to the infinite size is estimated to be
$\Delta_s\simeq 0.25J_0$ at $J_L=0.2$ while $\Delta_s \simeq 0.05 J_0$ at 
$J_L=0.3$.  In comparison with experimental conditions, we note that
the experimentally observed gap $\Delta_s\sim 500$K in
SrCu$_2$O$_3$\cite{17} with  a reasonable estimate $J_0\simeq 2000$K as deduced from the
susceptibility of similar but chain-type compounds
Sr$_2$CuO$_3$ and SrCuO$_2$\cite{18} leads to a similar ratio of 
$\Delta_s/J_0$ to
the case at $J_L=0.2$.  At $J_L=0.3$, $\langle M_s\rangle$ appears to 
follow a
power-law dependence $\langle M_s\rangle\sim ax^{q}$ with $q\sim 0.26$ 
for smaller
$x$ 
as predicted in the SC-MD
region of 2D systems.  This is reasonable because $J_L=0.3$ is close
to the  critical point $J_{Lc}$.  
The prefactor $b$ in the scaling of
the crossover boundary defined by  
$x_c\sim b \delta^{1.4}$
is roughly estimated as $b\sim0.5$ because
at $J_L=0.2$,
$x_c\sim$0.02 as we see in \ref{fig3}.     
Although the numerical results are rather 
preliminary, the results are
not inconsistent with the scaling prediction.

We next discuss the relevance of the SC-MD region for understanding
the experimental results.  Because the 3D coupling is
crucial to realize a nonzero $T_N$ and the 3D region dominates over the Q1D
and 2D regions for small $x$ as discussed above, the scaling properties in
experimental results for small $x$ need to be analyzed by the 3D
scaling.  
Although it is difficult to
estimate the exponents quantitatively from the available experimental
data, at 
least, it is consistent with  a quick increase of
$T_N$ as a function of $x$ with a wide SC region where even $x=0.01$
is involved.  Based on this qualitatively promising implication, further 
detailed comparisons of experimental results and
theoretical predictions here would be interesting.  
For example, clarification of pressure effects at both $x=0$ and
$x\not= 0$ would be desired because 
rather remarkable effects are expected through the control of $\delta$
near the QCP on the basis of the above theoretical analysis.  
In case of CaV$_2$O$_5$, it has been reported that dilute substitution
of Ti for V does not make the system antiferromagnetic\cite{19}.  
This difference may be understood by assuming much smaller
interladder exchange coupling than in the Cu compound due to presumable 
orbital ordering 
of $t_{2g}$ electrons on the V site.  

We also note that the closeness to the QCP 
has an important implication for  
the mechanism of superconductivity under doping of itinerant holes\cite{20}.
Here, we discuss two proposals for the mechanism.  One is that first
proposed by one of the authors\cite{21} where the carrier doping into
a spin-gapped Mott insulator may directly result in a superconducting
state in the region of the gap persistence.  This proposal 
was directly applied in the case of ladder models\cite{4,22}. 
  The other is the mechanism proposed
for the high-$T_c$ cuprates, where recovery of the suppressed
coherence inherent in 2D (or 3D) systems near the Mott transition 
results in the
superconductivity\cite{23,24}.  In this mechanism, the closeness to the
 QCP is
crucial.  The nonmagnetic impurity effects have turned out to be a
useful probe to measure the distance from the QCP
of the magnetic order.  The experimental results appear to support the
relevance of the quantum critical region and hence
support the latter mechanism.  

In summary, effects of interladder coupling on the antiferromagnetic 
transition have been investigated 
for coupled ladder systems
doped with nonmagnetic impurities.  Critical properties follow different 
scalings depending on four different regions.  They are separated by a 
crossover from the
weak-coupling to the strong-coupling  regions, as well as by a 
crossover from the quasi-one-dimensional to the
multi-dimensional regions.  The strong-coupling
and three-dimensional scaling behavior governed by the quantum
critical point of the undoped system dominates over a wide region,
which leads to characteristic power-law dependences of physical
quantities such as the N\'eel temperature on $x$.  We propose that
the scaling properties in this SC-3D region are relevant and important 
for understanding recent experimental indications in ladder compounds.

We thank H. Takagi, N. Furukawa, N. Katoh and Y. Motome
for useful
discussions.  This work is supported by a Grant-in-Aid for Scientific
Research on Priority Area ``Anomalous Metallic States near the Mott
Transition''from the Ministry of Education, Science and Culture.
The numerical calculations were performed at the
Supercomputer Center of the Institute for Solid State Physics, University
of Tokyo.


\end{document}